# Modeling the Line-of-Sight Integrated Emission in the Corona: Implications for Coronal Heating

Nicholeen M. Viall and James A. Klimchuk
NASA Goddard Space Flight Center

**Abstract**

One of the outstanding problems in all of space science is uncovering how the solar corona is heated to temperatures greater than 1 MK. Though studied for decades, one of the major difficulties in solving this problem has been unraveling the line-of-sight (LOS) effects in the observations. The corona is optically thin, so a single pixel measures counts from an indeterminate number (perhaps tens of thousands) of independently heated flux tubes, all along that pixel's LOS. In this paper we model the emission in individual pixels imaging the active region corona in the Extreme Ultraviolet. If LOS effects are not properly taken into account, erroneous conclusions regarding both coronal heating and coronal dynamics may be reached. We model the corona as a LOS integration of many thousands of completely independently heated flux tubes. We demonstrate that despite the superposition of randomly heated flux tubes, nanoflares leave distinct signatures in light curves observed with multi-wavelength and high time cadence data, such as those data taken with the Atmospheric Imaging Assembly on board the Solar Dynamics Observatory. These signatures are readily detected with the time-lag analysis technique of Viall & Klimchuk (2012). Steady coronal heating leaves a different and equally distinct signature that is also revealed by the technique.

## I. Introduction

A large body of research has taken shape over the last several decades on the subject of understanding the heating of the solar and stellar coronae. This 'coronal heating problem' is one of the major unresolved issues in space science, remaining a challenge despite the abundance of data, theory and modeling on the subject. A significant complication is the optically thin nature of the corona and the inherent spatial averaging in any observation. Even with spatially resolved solar observations, the detected emission is an integration of all of the flux tubes along the line-of-sight (LOS). One of the standard approaches to handling this LOS complication has been to analyze isolated features, such as coronal loops in active regions (AR). Coronal loops appear to follow the magnetic field and are identified and defined by their relative brightness as compared to the pixels next to them. A typical analysis approach is to estimate and subtract off the background emission (i.e. the emission due to flux tubes in front of and behind the loop), and then analyze the emission due to the loop itself (e.g. Klimchuk et al. 1992; Schmeltz et al. 2001, 2011; Warren et al. 2002, 2008; Del Zanna & Mason 2003, Winebarger et al. 2003; Winebarger & Warren 2005; Ugarte-Urra et al. 2006, 2009; Tripathi et al. 2009; Reale 2010; Aschwanden & Boerner 2011).

Much has been learned though these types of loop studies. For example, loop emission is dynamic and loops are typically found to be in a state of cooling (Winebarger & Warren 2005; Ugarte-Urra et al. 2009; Mulu-Moore et al. 2011; Viall & Klimchuk 2011, henceforth VK2011). They have been successfully modeled as bundles of subresolution, impulsively heated flux tubes

(Cargill & Klimchuk 1997; Warren et al. 2002, 2003; Winebarger et al. 2003; Winebarger & Warren 2005; Klimchuk 2006, 2009; Patsourakos & Klimchuk 2006, Bradshaw & Cargill 2006, 2010; Klimchuk et al. 2008). Due to the impulsive nature of the heating, observations do not show a progression of the loop from cool-to-hot during the heating phase followed later by a hot-to-cool progression during the cooling phase. Rather, the loop is observed first in a hot (>3 MK, e.g. AIA 335 Å channel) temperature observation, as the heating phase itself produces very little emission: the loop cools emitting in successively cooler channels at later times.

Though loop behavior and its properties are well understood and explained, loop emission is only somewhat brighter (10-30%) than the neighboring diffuse emission in between loops (Del Zanna & Mason 2003; VK2011). The large majority of the coronal plasma is in the diffuse emission, defined here as all coronal emission not contained within observationally distinct loops. This major component of the coronal plasma comprises the emission between loops and the 'background' emission along lines of sight containing loops. Both VK2011 and Viall & Klimchuk (2012) (henceforth VK2012) show diffuse locations where the light curves rise and fall in intensity with the peak intensity reached at later times in cooler channels, suggesting impulsive heating followed by cooling, even though no loops are present. This raises the question of whether the majority of the diffuse component is heated in a fundamentally different way or if it too is heated impulsively on small scales as loops are.

Recently VK2012 demonstrated a new technique to probe the properties of the spatially unresolved plasma along any LOS (including disk-integrated stellar and solar observations of spectral irradiance). It is an automated time-lag technique that identifies cooling patterns in data such as those from the Atmospheric Imaging Assembly (AIA) on the Solar Dynamics Observatory (SDO). In contrast to the standard loop analyses, this technique does not require user or automated identification of features (i.e. loops), or any background subtraction, all of which can introduce errors and biases. VK2012 present the results of using their technique on an AR and find clear evidence of at least some cooling plasma along the LOS in the majority of the pixels. Though their time-lag technique identifies the cooling patterns known to occur in loops, VK2012 find similar cooling patterns in diffuse emission locations where no discernible loops are present.

VK2012 demonstrated that the cooling patterns are qualitatively consistent with models of impulsive nanoflare heating. It is easy to see how a single coronal loop along the LOS would give rise to a coherent time-lag signature to be identified in the test. The diffuse emission is seemingly more difficult to understand. We propose that the diffuse emission in the AR is composed of flux tubes that are impulsively heated, just as loops are. However, there are up to tens of thousands along a given LOS, and they all are heated independently. We demonstrate that this scenario can produce an approximately steady composite light curve, one that in the absence of other information could be consistent with steady emission from steady heating. We show that steady heating in fact looks fundamentally different. We show that thousands of out-of-synch nanoflare light curves yielding one composite light curve still produces a coherent time-lag signal, and that the methods of VK2012 can distinguish between this scenario and one of truly steady emission from steady heating.

## II. Methods

The technique of VK2012 tests for the systematic cooling behavior known to occur in loop light curves. However, it tests every pixel and corresponding set of light curves in exactly the same way, whether there is a discernible loop present or not. They found widespread evidence for cooling plasma in the majority of the AR, consistent with impulsive heating. The technique developed and presented in VK2012 is as follows. They identify an area of interest (e.g. in VK2012 they chose NOAA AR 11082 on 19 June 2010) and create a subfield ('cutout') of that area. They derotate and coalign 24 hrs of continuous SDO/AIA data in the 335 (2.5 MK), 211 (2 MK), 193 (1.5 MK), 94 (1 and 7 MK), 171 (0.8 MK) and 131 (0.5 MK) Å EUV channels (see instrument response functions in Lemen et al. 2011; Boerner et al. 2011). At each pixel they create intensity time series, or light curves for each of the 6 channels. They compute the cross correlation value between two light curves (for example, the 211 and 193 Å light curves). Then, they recompute the cross correlation value after shifting the time series temporally relative to each other, recomputing the cross correlation value at every possible temporal offset up to 2 hrs forwards and backwards in time. They performed the same analysis on two consecutive twelve-hour time series, and twelve consecutive two-hour time series to test for persistence.

As shown in VK2012, the cross correlation value peaks at a given temporal offset and falls to low (uncorrelated) values at smaller and larger offsets. The time offset where the cross correlation value is greatest is the 'time-lag' between that particular pair of channels in that pixel. This time-lag is equivalent to the physical time it takes the plasma to evolve from the temperature of the first channel to the temperature of the second channel. Positive time-lags indicate cooling, due to the fact that VK2012 adopted the convention of placing the cooler channel second in the cross correlation lag.

The exception is pairs with the 94 Å channel: due to the bimodal temperature sensitivity of the 94 channel it may behave as either a warm 1 MK channel or a hot 7 MK channel depending on the differential emission measure distribution of the plasma along the LOS. VK2012 found that 94 Å usually behaved as a 'warm' channel except in the very central core of the AR where they found that hot plasma dominated the 94 Å light curves. Importantly, they found that all of the channel pairs, including those with 94 Å, were consistent with plasmas heated impulsively, reaching temperatures of greater than 3 MK (and in some cases, as hot as 7 MK) and then cooling to less than 1 MK. In addition to the finding that most of the pixels in the AR exhibit cooling time-lags, VK2012 also found that channel pairs whose peak temperatures are further apart have longer time-lags, consistent with the plasma cooling fully before any subsequent reheating.

Despite the knowledge gained by VK2012, all that they could conclude is that at least some of the plasma along the LOS must be cooling, rather than in a steady state at a constant temperature. The actual situation is likely very complex: the single time-lag measurement at a particular pixel is a composite time-lag, resulting from the behavior of many hundreds to perhaps tens of thousands of flux tubes along the LOS, all contributing to the emission and light curves. Here we introduce a modeling technique that attempts to reproduce this composite light curve observed in an AR pixel. We use the hydrodynamic simulation code called Enthalpy-Based Thermal Evolution of Loops (EBTEL) (Klimchuk et al. 2008) to simulate the emission from each individual flux tube. Because the hydrodynamic approach has primarily been applied to coronal loops, it is often referred to as "loop modeling," though we demonstrate how to expand the approach and apply it to the diffuse corona. It has been used by several authors to study a number of problems, ranging from soft X-ray loops (Lopez Fuentes & Klimchuk 2010) to active region arcades (Patsourakos & Klimchuk 2008), to flares (e.g., Raftery et al. 2009; Reeves & Moats 2010; Hock et al. 2013; Qiu et al. 2013).

EBTEL computes the evolution of the average temperature, density, and pressure along the coronal portion of a flux tube, or loop strand. Averaging along the tube is reasonable since thermal conduction and flows tend to smooth out any field-aligned gradients. The updated version of EBTEL used here has added improvements to the original code to account for gravitational effects and to more accurately treat the late phase of cooling (Cargill, Bradshaw, & Klimchuk 2012). EBTEL provides an excellent approximation to more sophisticated 1D codes, but using $10^3$-$10^4$ times less computing time. This is crucial for the type of model we show in this paper where an enormous number of simulations must be performed.

### III. Modeling the Diffuse Emission

The time-lags observed in coronal loops and those observed in the majority of the pixels in the AR analyzed by VK2012 can all be qualitatively explained by a single impulsively heated flux tube and by flux tubes that behave collectively (e.g. a nanoflare storm of finite duration; Klimchuk 2009). We show next that even a much larger number of nanoflare heated flux tubes that are completely uncorrelated will also produce a similar time-lag signature, detectable with the technique of VK2012. The time-lag signature is strong even when the resulting intensity fluctuations are small. This is one plausible physical scenario for the diffuse corona.

We use EBTEL to model the full LOS emission expected in a single pixel at the center of an AR as observed by SDO/AIA. This scenario includes many thousands to tens of thousands of emitting flux tubes which are heated stochastically. We assume low frequency nanoflares, meaning that each flux tube cools fully before any subsequent reheating. Some of the flux tubes may be part of an observational 'loop' due to random clusters of nanoflare events, but all of the flux tubes are heated entirely independently from one another and no nanoflare storms are explicitly input into the model. The basics of the model are as follows. First, we built a library of 1000 impulsively heated EBTEL simulations. For each simulation we construct light curves in each of the 6 AIA channels. We chose strand lengths corresponding to a range of field line lengths expected for a LOS through the core of AR 11082 (3-10.6 x $10^9$ cm half-lengths). The magnitude of the nanoflare (total energy per unit volume) is inversely dependent on the flux tube length squared (Mandrini, Demoulin, & Klimchuk 2000). For a given loop length, we simulate 10 possible nanoflare magnitudes, equal to 1-5 times the smallest magnitude, and 5 possible nanoflare durations between 50 and 250 seconds.

We assume that the nanoflares are stochastic, which we implement in the model by choosing a number of simulations from the library to begin every 1 second, the time step of the EBTEL runs. Both the number and choice of simulation are chosen randomly. At every time step we add together the light curves from all of the individual nanoflares to produce a composite light curve in each of the 6 SDO/AIA channels, plotted in Figure 1. We adjust the average rate of nanoflare occurrence so that the average heating rate along the LOS is consistent with the typical observed radiative loss rate of ARs. In the instance shown in Figure 1 the average rate of LOS nanoflare occurrence was 20 nanoflares per second. Since the process is random, sometimes many more nanoflares than this occur, sometimes far fewer.

We plot the predicted light curves for the 94 (red), 335 (green), 211 (blue), 193 (orange), 171 (cyan) and 131 (black) Å channels in Figure 1, each normalized to their own maximum and offset by -0.1 to avoid overlap. We use the updated (keyword /chiantifix) response functions to compute the light curves in this Figure; the conclusion that we draw in this paper are not dependent on the choice of response functions. In this case the light curves are 6 hours in length, but it is trivial to make them shorter or longer. With the parameters we have chosen here, approximately 70,000 flux tubes are contributing a non-zero emission to the pixel at any given time. Since there are so many flux tubes contributing emission to the pixel, the relative intensity change due to a single flux tube or nanoflare, is small.

The final time series has far smaller intensity variations than a background-subtracted loop light curve does. The difference between the maximum and minimum intensities, ΔI, is generally very small in each channel. It is the greatest in the 94 light curve, reaching 9% of the maximum. It is the smallest in the 335 Å light curve, where ΔI is a mere 5% of the maximum (the 335 Å light curve has the least variation of the 6 channels primarily due to the broader nature of that channel's response function). This level of predicted intensity fluctuation is even smaller than the level often observed in hot AR cores. For example, Warren et al. (2010) found that the intensity fluctuation were ~15% in single pixels imaging an AR core, which they provide as evidence of steady heating. In an actual observation this situation may be seemingly indistinguishable from steady emission from steady heating wherein the fluctuations are due to photon counting noise. However, the time-lag test used in VK2012 can disambiguate.

In Figure 2, we show the results of this time-lag test performed on our model light curves. We plot the cross correlation value as a function of temporal offset for five channel pairs: 211-193 Å (green), 335-211 Å (blue), 335-193 Å (orange), 335-171 Å (cyan) and 94-335 Å (red). Despite the relatively small variations that each strand contributes to the total light curves, the cross correlations exhibit the cooling patterns observed by VK2012. Namely, 211-193 Å, 335-211 Å, 335-193 Å, and 335-171 Å all reach a peak cross correlation value at a positive temporal offset. 211-193 has the shortest time-lag, while 335-171 Å exhibits the longest time-lag, consistent with plasma cooling fully from high to low temperatures. The hotter component dominates the 94 Å light curve in this particular case, so the positive 94-335 Å time-lag peak is large and evidence of a secondary negative peak due to the cooler component of 94 Å is not present. This is consistent with the positive 94-335 Å time-lag measured in the core of VK2012. As shown in VK2011, whether the hot component dominates the 94 Å light curves is dependent on the nanoflare strength. Based on this we conclude that the modeled positive 94-335 Å time-lag is a result of the strong nanoflares in our composite models.

The time lags and associated cross correlation values that we compute are summarized in Table 1. The channel pair with the shortest time lag and the strongest cross correlation is 211-193 Å while 335-171 Å channel pair exhibits both the longest time lag and the lowest cross correlation. It is significant that the numerical values of both the peak cross correlations and associated time-lags are generally similar in the model and diffuse pixel observations. For example, VK2012 found that most pixels exhibited a time-lag between 211-193 Å of less than ~ 600 seconds and cross correlation values of around 0.90. Detailed differences contain useful physical information. Although each nanoflare produces only a small change in intensity, each one follows the same hot-to-cool progression, so the composite intensity time series are still highly correlated at their respective cooling-time offsets. In fact, in the absence of noise, the amplitude of the intensity variations does not affect the cross correlation value, provided that the variations all follow the same hot-to-cool progression.

This model, though simple, demonstrates two important conclusions. First, light curves that are observed to be approximately steady are not necessarily produced by steady plasma. Though steadily heated flux tubes undergoing no appreciable evolution (on these sorts of time scales) is certainly a plausible explanation for an observation of a steady light curve, impulsive nanoflare heating is also a plausible explanation, provided there are many out-of-phase flux tubes along the LOS. Second, flux tubes with nanoflares that are completely stochastic and incoherent (i.e. never acting in a nanoflare storm to create a 'loop') and are still detectable with the time-lag analysis of VK2012.

## IV. Steady Heating and Noise

Some researchers have suggested that the diffuse emission in the cores of ARs is nearly constant as a result of effectively steady heating (Warren et al. 2010, 2011, 2012; Winebarger et al. 2011; though see Tripathi, Klimchuk, & Mason 2011; Schmelz & Pathak 2012; Bradshaw et al. 2012). Next we consider the case where 100% of the flux tubes along a LOS are steadily heated, producing steady emission. In this example we chose a representative pixel in the core of AR 11082 and use its observed count rates as the level of steady emission: 49 cts/s in 131 Å, 950 cts/s in 171 Å, 2481 cts/s in 193 Å, 1000 cts/s in 211 Å, 136 cts/s in 335 Å, and 12 cts/s in 94 Å. The resulting light curves are shown in grey in Figure 3 and are, by definition perfectly steady. We show the expected light curves when photon noise is added, assuming Poisson counting statistics, using the same colors as in Figure 1.

Figure 4 shows the cross correlation value computed between the same selected pairs of light curves in the same colors as shown in Figure 2. These cross correlation curves are fundamentally different from those of the observations (Figure 4 in VK2012) and those from the impulsive model (Figure 2). The cross correlation value randomly fluctuates, and is almost zero everywhere. None of the light curves are correlated at any temporal offset. This result is expected since the variability is due entirely to noise. For this particular example, the peak cross correlation value identified as the 'time-lag' happens to be -3049 s for 211-193, with a cross correlation value of 0.029, however a new generation of noise would equally likely result in any of the other tested time-lags, also at a very low cross correlation value. In fact, this result is independent of both the absolute noise level and the relative signal-to-noise level.

The other time-lag calculations show equally unphysical results at very low cross correlation values: we compute a time-lag of 774 s for 335-211 Å with a cross correlation value of 0.026; 200 s and a cross correlation value of 0.024 for 335-193 Å; -4688 s and a cross correlation value of 0.024 for 335-171 Å, and -646 and a cross correlation value of 0.023. Another indication of noise as opposed to true physical variability is that the greatest 'time-lag' and smallest 'time-lag' are not between the channel pairs with the greatest and smallest peak temperature difference. This random nature is the behavior of purely steady emission and noise, and will always result in an effectively zero (maximum of ~ 0.03) correlation, regardless of the count rate. VK2012 observed very few instances with such low cross correlation values between any of the channel pairs, so we can rule out the possibility that all of the plasma along the LOS is steady for the majority of the AR.

## V. Discussion and Conclusion

The basic model of part III has envisioned that all heating is impulsive and that the plasma in a given flux tube cools fully before the next nanoflare (low frequency heating). This is not the only possibility. Other scenarios include impulsive coronal heating with only partial cooling before subsequent reheating (high frequency nanoflares; e.g., Reep et al. 2013), thermal nonequilibrium (Mok et al. 2008; though see Klimchuk, Karpen, & Antiochos 2010) and hot coronal plasma coming from the tips of Type II spicules (De Pontieu et al. 2011; though see Klimchuk 2013). Another possibility is that some of the flux tubes along the LOS are heated steadily, undergoing steady emission as in part IV, but some of the flux tubes along the LOS are impulsively heated, as described in part III. An additional possibility is that of steady heating which lasts long enough for equilibrium to be established, but ends after some time, resulting in a cooling flux tube. We have begun to investigate models of these last two types, wherein there is a mix of both steadily heated flux tubes as well as impulsively heated flux tubes.

It is often assumed that the observation of approximately steady light curves indicates that the plasma creating those light curves must be steady. We have demonstrated with our model that impulsive nanoflare heating where the plasma is heated to hot temperatures and cools fully before subsequent reheating can also produce light curves that are approximately steady. It is possible for many thousands of impulsively heated strands that are all out of phase with one another, to produce a LOS integrated light curve that is remarkably steady. The $\Delta I$ peak-to-trough variability in a light curve is not an indication of the amount of emission due to impulsively heated flux tubes along the LOS; it is merely a lower limit.

Importantly, we demonstrated in this paper that even seemingly steady light curves produced through thousands of out-of-phase nanoflare heated flux tubes still retain the cooling signatures from the nanoflares, which the time-lag test of VK2012 will identify. Lastly, we demonstrated that purely steady emission as a result of steady heating will produce a 'time-lag' that is not physically meaningful, and a cross correlation value of no more than about 0.03 since the variability is entirely due to noise. In any observed instances with cross correlation value significantly greater than ~ 0.03 we can immediately rule out a scenario in which all of the flux tubes along the LOS are heated steadily producing steady emission. For VK2012 this means that we can rule out steady emission resulting from entirely steady heating for the vast majority of the AR.

**Acknowledgements**

This research was supported by the NASA Supporting Research and Technology program. We thank the reviewer for their helpful comments and suggestions. The data are courtesy of NASA/SDO and the AIA science team. This work benefited greatly from the International Space Science Institute team meeting **'**Coronal Heating – Using Observables to Settle the Question of Steady vs. Impulsive Heating' led by Stephen Bradshaw and Helen Mason.

# Table 1. Time Lags and Cross Correlation Values

| Channel Pair | Time Lag (seconds) | Cross Correlation Value |
|---|---|---|
| 211-193 Å | 146 | 0.96 |
| 335-211 Å | 488 | 0.83 |
| 335-193 Å | 676 | 0.72 |
| 94-335 Å | 740 | 0.76 |
| 335-171 Å | 900 | 0.64 |

**Table 1.** Summary of time lags (in seconds) and cross correlation values calculated between the pairs of modeled light curves. Corresponds to cross correlations shown in Figure 2.

**Figure Captions**
**Figure 1** SDO/AIA light curves predicted with EBTEL nanoflare simulations of diffuse AR core emission. 94 shown in red, 335 in green, 211 in blue, 193 in orange, 171 in cyan, and 131 in black; each is normalized to its maximum and offset by -0.1. Light curves are remarkably steady even though all flux tubes are heated impulsively.

**Figure 2** Cross correlation as a function of temporal offset for pairs of the modeled diffuse emission light curves shown in Figure 1. Time-lags occur where the cross correlation peaks, indicated with dots. Green represents the 211-193 pair, blue the 335-211 pair, orange the 335-193 pair, cyan the 335-171 pair, and red the 335-94 pair.

**Figure 3** Light curves for steady emission in the different AIA channels shown as grey horizontal lines and with the addition of photon noise (same colors as in Figure 1): the light curves expected resulting from steady heating.

**Figure 4** Cross correlation as a function of temporal offset computed between pairs of steady heating light curves shown in Figure 3. The curves are mostly hidden behind the green (211-193) curve. Time-lags (indicated with dots) are random and occur at an extremely low cross correlation value.

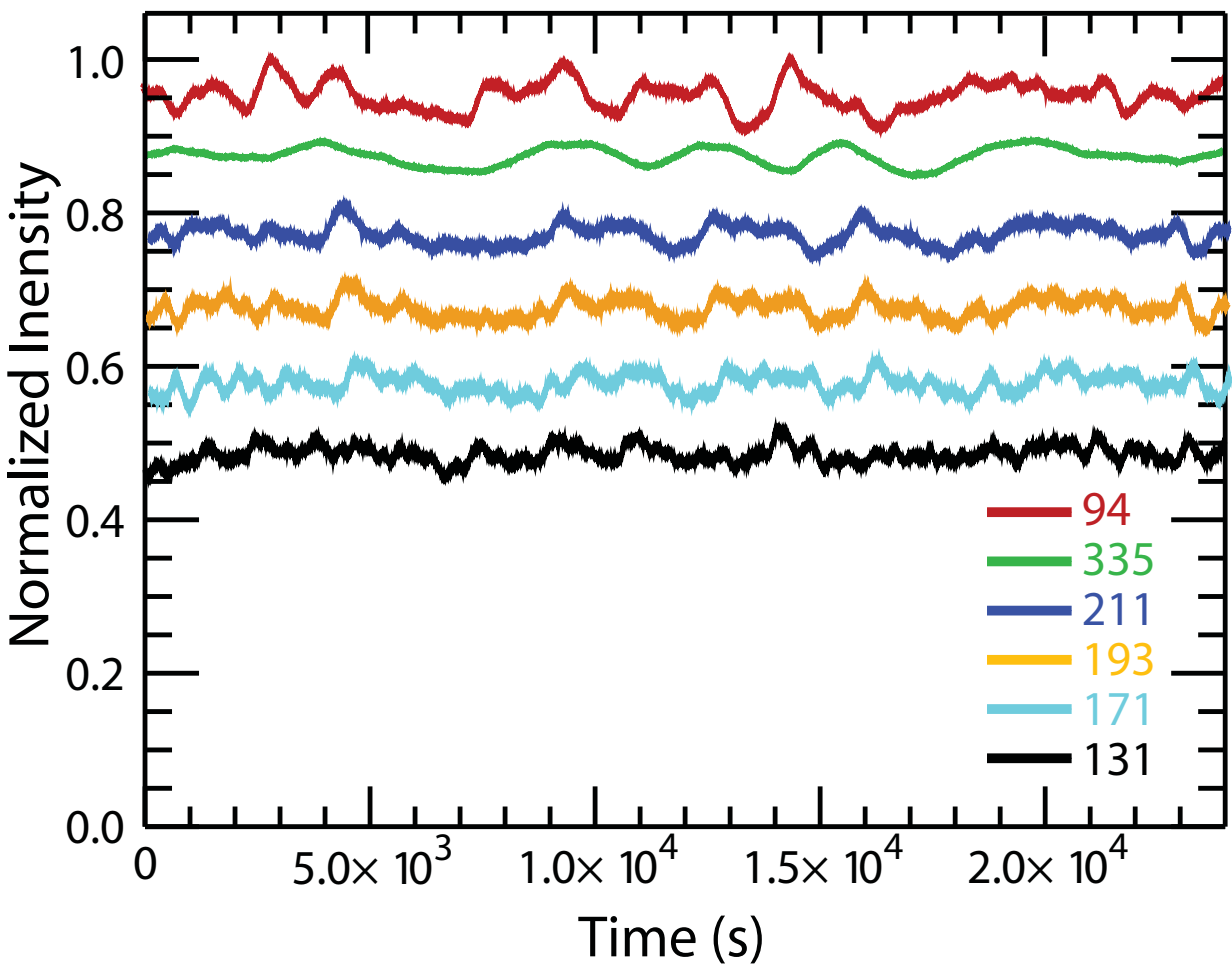

Figure 2

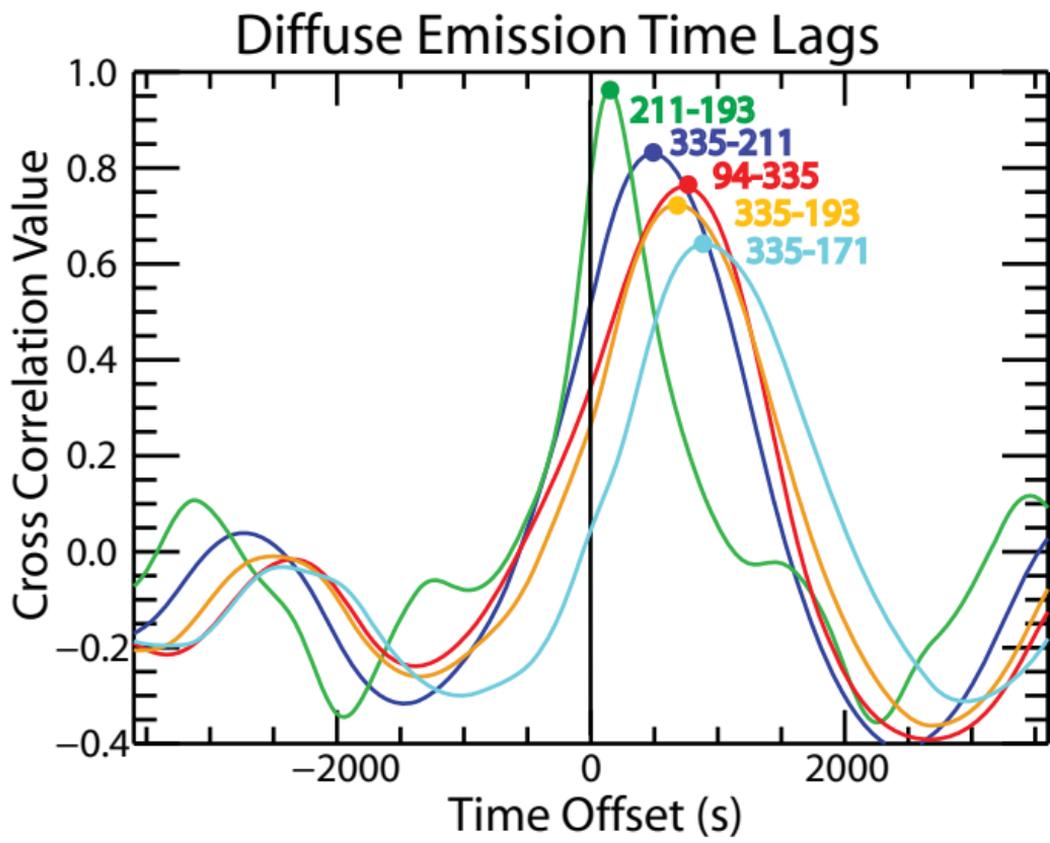

Figure 3

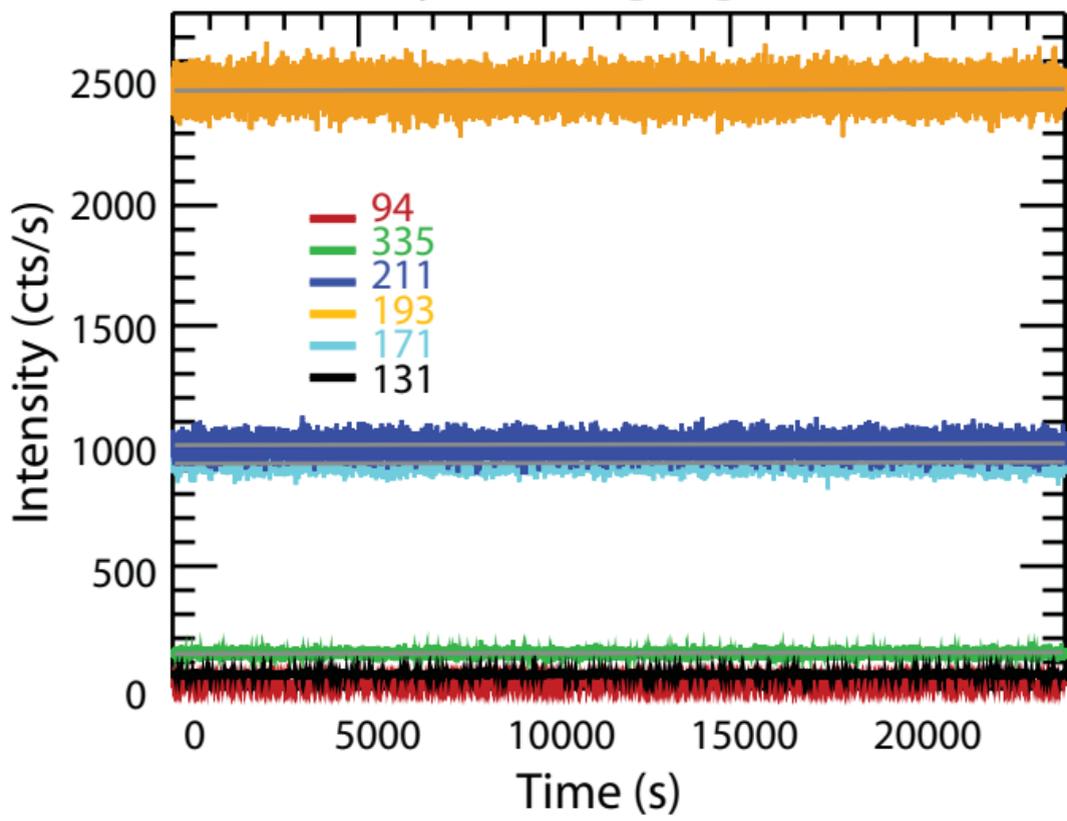

Figure 4

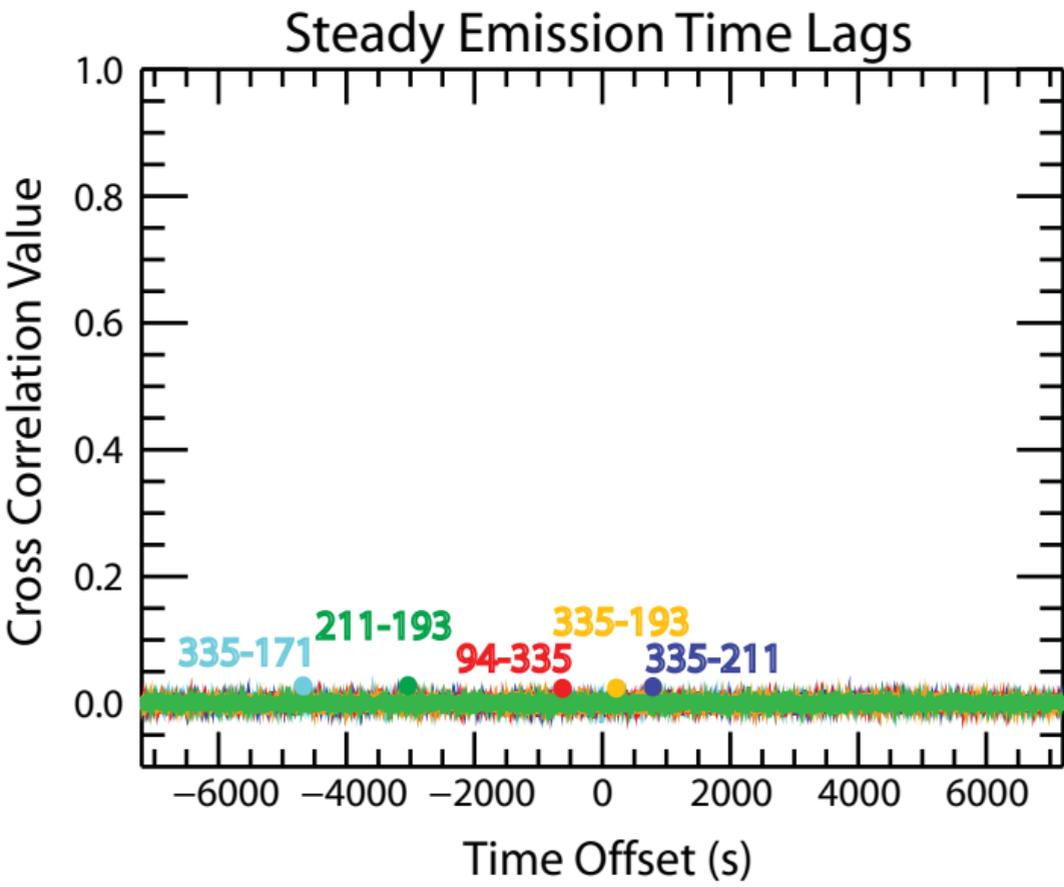